\documentclass[fleqn,usenatbib]{mnras}

\usepackage{newtxtext,newtxmath}
\usepackage[T1]{fontenc}
\DeclareRobustCommand{\VAN}[3]{#2}
\let\VANthebibliography\thebibliography
\def\thebibliography{\DeclareRobustCommand{\VAN}[3]{##3}\VANthebibliography}
\usepackage[utf8]{inputenc}
\usepackage{natbib}
\usepackage{color}
\usepackage{subfigure}
\usepackage{physics}
\usepackage{mathrsfs}
\usepackage{amsmath}

\usepackage{dcolumn}% Align table columns on decimal point
\usepackage{hyperref}
\usepackage{lineno}
\usepackage{graphicx} % Required for inserting images
\usepackage[normalem]{ulem}

\definecolor{pink}{RGB}{255, 20, 147}     

\definecolor{applegreen}{rgb}{0.55, 0.71, 0.0}

\definecolor{carminered}{rgb}{1.0, 0.0, 0.22}

\definecolor{purple}{rgb}{0.533, 0.0, 1.0}

\definecolor{blue-violet}{rgb}{0.30, 0.1, 0.89}

\definecolor{verde}{rgb}{0.01, 0.53, 0.31}

\title[APTA: a new dHz GW detector concept]{Artificial Precision Timing Array: bridging the decihertz gravitational-wave sensitivity gap with clock satellites}

\author[L. M. B. Alves et al.]{
Lucas M. B. Alves,$^{1,2}$\thanks{E-mail: lucas.alves@columbia.edu}
Andrew G. Sullivan,$^{3}$
Xingyu Ji,$^{1}$
Doğa Veske,$^{4,5,6}$
Imre Bartos,$^{7}$
Sebastian Will,$^{1}$
\newauthor
Zsuzsa Márka,$^{5}$
and Szabolcs Márka$^{1}$
\\
$^{1}$Department of Physics, Columbia University in the City of New York, New York, NY 10027, USA\\
$^{2}$Departamento de Física, Universidade Federal de Minas Gerais, Belo Horizonte, MG 31270-901, Brazil\\
$^{3}$Kavli Institute for Particle Astrophysics and Cosmology, Department of Physics, Stanford University, Stanford, CA 94305, USA\\
$^{4}$Institut für Theoretische Physik, Universität Heidelberg, Heidelberg 69120, Germany\\
$^{5}$Columbia Astrophysics Laboratory, Columbia University in the City of New York, New York, NY 10027, USA\\
$^{6}$Fizik Bölümü, Orta Doğu Teknik Üniversitesi, Çankaya/Ankara 06800, Turkey\\
$^{7}$Department of Physics, University of Florida, PO Box 118440, Gainesville, FL 32611, USA
}

\date{Accepted XXX. Received YYY; in original form ZZZ}

\pubyear{2025}

\begin{document}
\label{firstpage}
\pagerange{\pageref{firstpage}--\pageref{lastpage}}
\maketitle

\begin{abstract}
Gravitational-wave astronomy has developed enormously over the last decade, with the first detections and continuous development across broad frequency bands. However, the decihertz range has largely been left out of this development. Gravitational waves in this band are emitted by some of the most enigmatic sources, including intermediate-mass binary black hole mergers, early inspiraling compact binaries—whose mergers are seen by Earth-based detectors—, and possibly primordial gravitational waves. To tap this exciting band, we propose the construction of a detector based on pulsar timing principles, the Artificial Precision Timing Array (APTA). We envision APTA as a solar system array of artificial ``pulsars''—precision-time-reference-carrying satellites that emit periodic electromagnetic signals towards Earth or another satellite constellation receiver location. In this fundamental study, we estimate the clock precision needed for gravitational-wave detection with APTA. Our results suggest that 6 satellites and a clock relative uncertainty of $10^{-18}$ at 1~s of averaging, which is currently attainable with ground-based atomic clocks, would be sufficient for APTA to reach pristine sensitivity in the decihertz band and observe $10^3$–$10^4$ $\mathrm{M}_\odot$ black hole mergers and the early inspiral of heavy LIGO-Virgo-KAGRA sources. Future clock and oscillator technologies realistically expected in the next decade(s) would enable the detection of an increasingly diverse set of sources, allowing APTA to reach a better sensitivity than other detector concepts proposed for the decihertz band. This work opens up a new area of research into designing and constructing gravitational-wave detectors relying on principles used successfully in pulsar timing.
\end{abstract}


\section{Introduction}
Since the first detection of gravitational waves (GWs) by LIGO in 2015 \citep{LIGOScientific:2014pky,LIGOScientific:2016aoc,LIGOScientific:2021djp, LIGOScientific:2025slb}, more than 300 compact-binary mergers have been seen by ground-based detectors. Moreover, the global network of pulsar timing arrays (PTAs) has provided strong evidence for a stochastic GW background \cite{VIRGO:2012dcp,2019kagra,NANOGrav:2023gor, 2023EPTAdiscovery, 2023ParkesPTA, 2023Parkes_PTADR, 2023ChinesePTA}. With the fourth LIGO-Virgo-KAGRA observing run in progress and the continued PTA campaign \citep{2013PASA...30...17M,VIRGO:2012dcp,Antoniadis:2022pcn,NANOGrav:2023gor}, more and more observational data and, thus, scientific insight can be expected.

While Earth-based detectors utilize interferometry to detect GWs, PTAs, including NANOGrav \citep{NANOGrav:2023gor}, the European Pulsar Timing Array \citep{2016MNRAS.458.3341D, 2023EPTAdiscovery}, the Parkes Pulsar Timing Array \citep{2013PASA...30...17M, 2023ParkesPTA, 2023Parkes_PTADR}, the Chinese Pulsar Timing Array \citep{2023ChinesePTA}, and the MeerKAT Pulsar Timing Array \citep{2020PASA...37...28B}, rely on the pulsar timing technique. Pulsars, rapidly rotating magnetized neutron stars, emit regular radio pulses with periods ranging from $10^{-3}$ to 10~s  \citep{2016era..book.....C}. The change in their spin periods $\dot{P}$  is $\dot{P}\lesssim10^{-19}$~ss$^{-1}$ \citep{2008LRR....11....8L}. With these very precisely measured and stable periods over long timescales, millisecond pulsars make for some of the powerful natural clocks. A GW passing between a pulsar and the Earth alters the travel time of the radio pulse from the star to the telescope. Subtracting a model for the times of arrival (ToAs) from the observed data yields a set of timing residuals, deviations from the expected ToAs, which may come from GWs or noise processes. While optimized for stochastic GW detection, PTAs are capable of detecting continuous signals \citep{2015CQGra..32a5014M, Maggiore:2018sht}, and, in fact, NANOGrav has conducted continuous GW searches for supermassive BBH candidates \citep{2025arXiv250816534A}. Our focus in this work is on the application of pulsar timing principles to the detection of individual sources rather than stochastic backgrounds.

Current techniques access a limited window of the GW frequency spectrum. The ongoing PTA observation campaign is sensitive in the $10^{-9}$–$10^{-6}$~Hz band \citep{2021Symm...13.2418M}. Usually, PTAs have their upper-frequency limit set by their fortnightly access to radio telescopes. Even when this practical concern is overcome, there is a fundamental frequency upper limit of order 1~mHz due to the folding time required to process pulsar observations \citep{Dolch_2014,Dolch_2016}. Meanwhile, LIGO-Virgo-KAGRA interferometers are sensitive in the 10 Hz–10 kHz band. In the future, space-based interferometers such as the Laser Interferometer Space Antenna (LISA) and TianQin may be sensitive to $0.1$–$100$~mHz GWs, while next-generation Earth-based missions such as Cosmic Explorer (CE) and Einstein Telescope (ET) will push the lower limit of ground-based sensitivity to a few~Hz \citep{Luo_2016,LISA:2017pwj,2019BAAS...51g..35R,ETzao_cabul}. Still, seismic and gravity gradient noises severely limit ground-based detector sensitivity below $\sim$ 1~Hz \citep{LIGOScientific:2007fwp,Punturo:2010zz,LIGOScientific:2016aoc}. Consequently, the 0.1–1~Hz region remains inaccessible, although nice concepts, such as the Atom Interferometer Observatory and Network (AION) \citep{Badurina:2019hst}, the Advanced Laser Interferometer Antenna (ALIA) \citep{Bender:2004vw,bender_begelman_2005}, the Big Bang Observer (BBO) \citep{phinney2004big,Harry_2006}, the DECi-hertz Interferometer Gravitational wave Observatory (DECIGO) \citep{Kawamura:2006up}, the Gravitational-Wave Lunar Observatory for Cosmology (GLOC) \citep{2020arXiv200708550J}, and the Lunar Gravitational-wave Antenna (LGWA) \citep{ajith2024lunargravitationalwaveantennamission} have been proposed.

Observational access to the 0.1–10~Hz GW band is imperative for the progress of many areas of astrophysics and cosmology. For example, $10^3$–$10^4$ $\mathrm{M}_\odot$ binary black holes (BBHs) have yet to be unambiguously observed. The observation of these sources can advance our understanding of supermassive black hole and galaxy formation, stellar clusters, and active galactic nuclei (AGN) \citep{2004IJMPD..13....1M,2009ApJ...696L.146M,McKernan_2012,2014MNRAS.443.2410F,2019PhRvL.123r1101Y}. While LISA may detect the early inspirals of $10^3$–$10^4$ $\mathrm{M}_\odot$ binaries \citep{2020NatAs...4..260J}, a 0.1–10~Hz detector accesses the wealth of additional information only available during their mergers. Other promising phenomena that may feature in this band include primordial GWs \citep{Corbin_2006,Smith_2006,Kuroyanagi_2009,Guzzetti:2016mkm}, the evolution of stellar-mass binaries and triple black hole systems \citep{2021A&A...650A.189A,10.1111/j.1365-2966.2009.14653.x,2018PhRvD..97j3014S,2018ApJ...855...34G,2017ApJ...841...77A,10.1093/mnras/sty2334, 2023PhRvD.107b3023C}, supernovae \citep{2009CQGra..26f3001O,2015PhRvD..92l4013S}, and parabolic encounters \citep{2006ApJ...648..411K, 2023PhRvD.107b3023C}, among others further discussed in Section \ref{sec2}.

To reach this unexplored GW band, we propose a new detector concept: the Artificial Precision Timing Array (APTA), a network of satellites distributed around the Earth or another receiver carrying high-precision clocks or comparably accurate time references to act as artificial ``pulsars''. Having APTA clock satellite signals be clearly modulated and aimed at a small station (or stations) dedicated to their reception, APTA can avoid the cadence constraint of regular PTAs and achieve high sensitivity at much higher frequencies. If the satellite signals are pulsed at more than 10~Hz, APTA can reach high sensitivity to 0.1–10~Hz (the APTA band) GWs for observations lasting more than 10 s.

The idea of using high-precision clocks in space for the empirical study of GWs has been proposed and applied before. The global positioning system (GPS) was used to constrain the strain amplitude of the GW background in the 0.01–1~Hz band \citep{PhysRevD.89.067101}, in an application of the Doppler tracking method \citep{1975GReGr...6..439E,2006LRR.....9....1A,Vutha_2015}. Another study proposed the use of two satellites in heliocentric orbits equipped with optical lattice atomic clocks to detect GWs, using a technique similar to Doppler tracking \citep{PhysRevD.94.124043}. Another proposed detector, the Space Atomic Gravity Explorer (SAGE), consists of two satellites carrying ultracold atomic strontium systems to either perform atom interferometry with the strontium or Doppler tracking utilizing Sr optical atomic clocks \citep{2019EPJD...73..228T}. Nevertheless, the idea of performing pulsar-timing-type GW detection with artificial satellites equipped with high-precision clocks is unprecedented.

In this paper, we describe the groundwork for APTA targeting the 0.1–10~Hz GW frequency band. In Section \ref{sec2}, we list and discuss several interesting astrophysical targets observable in this band. In Section \ref{sec3}, we derive a characteristic strain sensitivity curve for APTA. In Section \ref{sec4}, we present the concept of APTA in detail, discussing parameter choices for the detector and showing its sensitivity curves alongside other detectors and sources. In Section \ref{sec5}, we briefly list and discuss error-introducing factors to be incorporated into a more realistic future model of APTA. We conclude in Section \ref{sec6}.

\section{Gravitational-Wave Astrophysics with APTA}
\label{sec2}
The GW frequency regime between LISA and Earth-based detectors probes a diverse set of astrophysical and cosmological sources \citep{2020CQGra..37u5011A}. Notably, $10^3$–$10^4$ $\mathrm{M}_\odot$ black hole binaries, which merge before entering the LIGO-Virgo-KAGRA band, remain rich for study. The nature of these sources provides unparalleled insight into the formation of supermassive black holes and galaxies as well as AGN and stellar cluster dynamics \citep{2004IJMPD..13....1M,2009ApJ...696L.146M,McKernan_2012,2014MNRAS.443.2410F}.

Inspiraling stellar-mass compact binaries also emit GWs in this band before merging in the ground-accessible band. Low-frequency GW observations of these sources directly probe their formation and evolution \citep{2016PhRvD..94f4020N,2017MNRAS.465.4375N, 2021A&A...650A.189A,10.1111/j.1365-2966.2009.14653.x,2018PhRvD..97j3014S,2018ApJ...855...34G,2017ApJ...841...77A}. Long-duration tracking may probe BBH eccentricity \citep{2020CQGra..37u5011A,2022PhRvD.105b3019L} as well as improve effective spin and localization measurements. Similarly, the formation of merging triple black hole systems may also be probed in this band \citep{10.1093/mnras/sty2334, 2023PhRvD.107b3023C}. Excitingly, early detection of an inspiraling compact binary could trigger a wide range of observations in multimessenger astronomy, such as counterpart and precursor searches \citep{ 2012PhRvL.108a1102T,    2023MNRAS.520.6173S}, as well as give advance warning for gravitational-wave detectors in higher frequency bands, gamma-ray-burst observatories \citep{1992ApJ...395L..83N, 2017ApJ...848L..12A, 2017ApJ...848L..13A}, and telescopes looking for kilonova or other multimessenger signatures (see \cite{gottlieb2023unifiedpictureshortlong} and references therein). 

The characteristic GW strain of an astrophysical binary at redshift $z$ may be estimated as \citep{Sesana_2005}
\begin{multline}
\label{eq:strain}
    h_c\approx 5.3\times10^{-20} \left(\frac{f_{\rm gw}}{0.1 \text{~Hz}}\right)^{-\frac{1}{6}}\left(\frac{M}{10^3 \text{ M}{_\odot}}\right)^{\frac{5}{6}} \frac{q^{\frac{1}{2}}}{(1+q)}\\ \left(\frac{(1+z)}{1.2}\right)^{\frac{5}{6}}\left(\frac{D}{1 \text{~Gpc}}\right)^{-1},
\end{multline}
where $M$ is the source-frame total mass, $f_{\rm gw}$ is the observed GW frequency, $q\leq1$ is the mass ratio, and $D$ is the luminosity distance to the source, which directly determines $z$ given a set of cosmological parameters—which we take from Planck 2018 for this study \citep{Planck:2018vyg}. The merger frequency for a BBH may be estimated with a Newtonian approximation as
\begin{equation}
\label{eq:mergefreq}
    f_{\rm gw}^{\rm merger}\approx\frac{1}{\pi}\frac{c^3}{2^\frac{3}{2}GM}\approx2.3\times10^4\text{~Hz} \left(\frac{M}{1 \text{ M}_\odot}\right)^{-1}.
\end{equation}
Heavy stellar-mass BBHs (total mass $\sim100$ $\mathrm{M}_\odot$) at distances of $\sim1$~Gpc will have strains of order $h_c\sim 10^{-21}$ in the dHz band during their early inspiral. The late inspirals and mergers of $10^3$–$10^4$ $\mathrm{M}_\odot$ BBHs should have even larger dHz GW strains. For instance, an intermediate-mass binary black hole merger with $D\sim1$~Gpc, and total mass $\sim1\times10^4$ M$_\odot$, which will merge at $f_{\rm gw}\approx2.3$~Hz, will have strains of $h_c\sim10^{-19}$. 

Since the GW frequency is twice the orbital frequency, the time in the detector frame over which massive merging BBHs will emit 0.1-10~Hz GWs may be estimated as \citep{PhysRev.136.B1224}
\begin{equation}
    \label{separation}
    T\approx 70 \text{~s }\left(\frac{1+z}{1.2}\right) \left(\frac{M}{10^4\text{ M}}_\odot\right)^{-\frac{5}{3}}\left(\frac{f_{\rm gw}}{0.1 \text{~Hz}}\right)^{-\frac{8}{3}} \frac{(1+q)^2}{q}.
\end{equation} Therefore, $10^3$–$10^4$ M$_\odot$ binaries will spend $10^2$–$10^4$~s in the APTA band before merging, while stellar-mass BBHs can remain for days to even years depending on the mass, allowing for long-term signal accumulation from these sources.

More unique astrophysical GW sources may also manifest in this frequency regime. Low-frequency sources such as Type Ia \citep{2015PhRvD..92l4013S} and core-collapse supernovae \citep{2009CQGra..26f3001O}, parabolic encounters \citep{2006ApJ...648..411K, 2023PhRvD.107b3023C}, white dwarf-neutron star mergers \citep{2019MNRAS.482.3656R,2022MNRAS.511.5936K,2024MNRAS.528.5309K}, and even some extreme-mass-ratio inspirals \citep{2000PhRvD..62l4021F, 2000PhRvD..61h4004H, 2017PhRvD..95j3012B} should emit GWs in this band. With longer observations, APTA can tap lower frequencies. As such, APTA could be able to observe verification binaries to be used in LISA calibration if their operational lifespans intersected \citep{2018MNRAS.480..302K,2011CQGra..28i4019M,2017MNRAS.470.1894K,2018ApJ...854L...1B,2018PhRvL.121m1105T}, as well as probe the major astrophysics goals of LISA \citep{2023LRR....26....2A}. Similarly, there will be major synergies with future ground-based detectors such as CE \citep{2019BAAS...51g..35R} and ET \citep{Punturo:2010zz}, as APTA will observe the very early inspirals of binaries which will merge in the ground-based band.

Observations in the $0.1$–$10$~Hz frequency band could also help constrain models of early-universe physics through the detection of primordial gravitational waves \citep{Corbin_2006,Smith_2006,Kuroyanagi_2009,Guzzetti:2016mkm}. It is advantageous to explore this frequency range as opposed to LISA's $0.1$–$100$~mHz \citep{LISA:2017pwj} since, in the latter, primordial GWs may be overshadowed by binary white dwarf GWs, which disappear entirely above 0.25~Hz \citep{PhysRevD.63.064030,10.1111/j.1365-2966.2003.07176.x,Cutler_2009}. Furthermore, roughly 10\% of the dark matter could be in the form of primordial black holes (PBHs) with masses of $\sim1$ $\mathrm{M}_\odot$ (see region C in Fig. 4 of \cite{Carr_2022}). Binaries of these PBHs could be detected in the APTA band during their early inspiral \citep{Nakamura_1997, Raidal_2019}. Another possibility of cosmological interest in the APTA band could be the direct measurement of the acceleration of the expansion of the universe and, therefore, the curvature parameter $\Omega_{k0}$ \citep{Nakamura_1999,Seto_2001}. 

\section{APTA Sensitivity}
\label{sec3}

\subsection{The Physics of Pulsar-Timing-Based GW Detection}
\label{sec3a}
In this subsection, we provide a theoretical overview of the physical mechanism underlying how PTAs, and hence APTA, work. Our calculation follows closely the one in \cite{Maggiore:2018sht} and we initially work in $c=1$ units.

Gravitational waves are linear perturbations to a flat background spacetime. In transverse-traceless (TT) gauge, suitable to study GW propagation in vacuum,  the spacetime line element $ds$ to linear order is defined by
\begin{equation}
    ds^2=-dt^2+ \left[ \delta_{ij}+h_{ij}^{TT}(t,\Vec{x}) \right] dx^idx^j.
\end{equation}
where $t$ is the time coordinate, $x^i$, $i \in \{1,2,3\}$, are spatial coordinates, $\delta_{ij}$ is the Kronecker delta and $h_{ij}^{TT}$ is the TT gravitational-wave tensor.

We consider a coordinate system with its origin at APTA's signal reception station, in which one of its satellites is at position $\Vec{r_s}=d_s\mathbf{\hat{x}}$ on the x-axis (the basic architecture of APTA is discussed in Section \ref{sec4}). Following the null (i.e. $ds^2=0$) trajectory $\Vec{x}(t)$ of a photon emitted at time $t_e$ from the satellite to the receiver, 
\begin{equation}
    dx^2=\frac{dt^2}{1+h^{TT}_{xx}(t,\Vec{x}(t))}.
\end{equation}
The photon is traveling from positive $x$ to the origin, so we get the solution moving in the $-x$ direction:
\begin{equation}
    dx=\frac{-dt}{\sqrt{1+h^{TT}_{xx}(t,\Vec{x}(t))}}\approx -\left[1-\frac{1}{2}h^{TT}_{xx}(t,\Vec{x}(t))\right]dt
\end{equation}
to first order in the perturbation $h_{xx}^{TT}$. 

When dealing with a single pulsar or APTA satellite, one can always choose the coordinate system so that the light source sits along one of the axes. However, to proceed to a result for a light source on an arbitrary direction $\mathbf{\hat p}$, which will be useful when we deal with multiple beacons, we replace $h_{xx}^{TT}$ with $\hat p^i \hat p^j h^{TT}_{ij}$.

Next, we can integrate from $t_e$ to the observation time $t_o$ so that
\begin{equation}
    d_s=t_o-t_e-\frac{\hat p^i \hat p^j}{2}\int_{t_e}^{t_o}\dd t' \;h^{TT}_{ij}(t',\Vec{x}(t')).
\end{equation}
We are concerned with first-order perturbations and the integrand $h^{TT}_{ij}$ is already a first-order term. Therefore, within the integral, we can consider the unperturbed trajectory of the photon,
\begin{subequations}
    \begin{align}
        &t_o=t_e+d_s \\
        &\Vec{x_0}(t)=(t_e-t+d_s)\mathbf{\hat{p}},
    \end{align}
\end{subequations}
such that
\begin{equation}
\label{timing1}
    t_o=t_e+d_s+\frac{\hat p^i \hat p^j}{2}\int_{t_e}^{t_e+d_s}\dd t' \;h^{TT}_{ij}(t',(t_e+d_s-t')\mathbf{\hat{p}}).
\end{equation}

The fundamental idea in pulsar timing is to detect gravitational waves through deviations in the time elapsed between observed pulses. Therefore, consider the immediate next pulse, emitted by the satellite at time $t_e+\tau$, where $\tau$ is the pulsing period of the light source. The second pulse will be observed at $t_o'$. The same derivation that led to Eq. \ref{timing1} gives, by replacing $t_o$ and $t_e$ with $t_o'$ and $t_e+\tau$,
\begin{equation}
\label{timing2}
    t_o'=t_e+\tau+d_s+\frac{\hat p^i \hat p^j}{2}\int_{t_e+\tau}^{t_e+\tau+d_s}\dd t' \;h^{TT}_{ij}(t',(t_e+\tau+d_s-t')\mathbf{\hat{p}}).
\end{equation}
Changing the integration variable in Eq. \ref{timing2} as $t'\rightarrow t'-\tau$ and subtracting Eq. \ref{timing1} from it, it becomes clear that the effect of the GW is a pulsing period shift as perceived by the observer: 
\begin{subequations}
\label{deltatau}
\begin{align}
    &t_o'-t_o=\tau+\Delta\tau \\
    &\Delta\tau=\frac{\hat p^i \hat p^j}{2}\int_{t_e}^{t_e+d_s}\dd t' \; \left[h^{TT}_{ij}(t'+\tau,\Vec{x_0}(t'))-h^{TT}_{ij}(t',\Vec{x_0}(t'))\right].
\end{align}
\end{subequations}

For the usual PTA, the typical source period $\tau_{p}$ is of order 1 ms while the target GW frequency $\omega_{gw,p}$ is of order $0.1\,\mathrm{yr^{-1}}$, such that $\omega_{gw,p}\tau_{p}\ll1$. The validity of this assumption for APTA is discussed in Section \ref{sec4a}. In this regime, we may Taylor expand $h^{TT}_{ij}(t'+\tau,\Vec{x_0}(t'))$ around $t'$ in powers of $\tau$ and discard second-order and higher terms:
\begin{equation}
    h^{TT}_{ij}(t'+\tau,\Vec{x_0}(t'))= h_{ij}^{TT}(t',\Vec{x_0}(t'))+\frac{\partial}{\partial t'}h_{ij}^{TT}(t',\Vec{x_0}(t'))\tau
\end{equation}
so that Eq. \ref{deltatau} becomes
\begin{equation}
\label{approx-integral}
    \frac{\Delta\tau}{\tau}=\frac{\hat p^i \hat p^j}{2}\int_{t_e}^{t_e+d_s}\dd t' \; \left[\frac{\partial}{\partial t'}h_{ij}^{TT}(t',\Vec{x})\right]_{\Vec{x}=\Vec{x_0}(t')}.
\end{equation}

To integrate Eq. \ref{approx-integral}, we now specialize to a monochromatic GW—which, depending on the observational timescale, may be a good approximation for several astrophysical sources—propagating along direction $\mathbf{\hat n}$ in TT gauge: 
\begin{equation}
    h_{ij}^{TT}(t,\Vec{x})=\mathcal{A}_{ij}(\mathbf{\hat{n}})\cos{\left[\omega_{\rm gw}(t-\mathbf{\hat{n}}\cdot \Vec{x})\right]},
\end{equation}
where the amplitude tensor satisfies the transverse condition $n^i\mathcal{A}_{ij}(\mathbf{\hat{n}})=0$. With this, Eq. \ref{approx-integral} becomes:
\begin{equation}
\label{pre-template}
    \frac{\Delta\tau}{\tau}=\frac{\hat p^i \hat p^j\mathcal{A}_{ij}}{2(1+\mathbf{\hat p}\cdot\mathbf{\hat n})}\{\cos[\omega_{\rm gw}t_o]-\cos[\omega_{\rm gw}(t_e-d_s\mathbf{\hat p}\cdot\mathbf{\hat n})]\}.
\end{equation}

PTA signal processing can be done with either the so-called redshift, expressed in Eq. \ref{pre-template}, or with its integral, the timing residual $R(t)$:
\begin{equation}   
    \label{template}
    \begin{split}
    R(t)  = \int_0^t   \dd t' \; \frac{\Delta \tau}{\tau}&(t') = \frac{\hat p^i \hat p^j\mathcal{A}_{ij}}  {2(1+\mathbf{\hat p}\cdot\mathbf{\hat n})} \frac{1}{\omega_{\rm gw}}  \Big\{\sin{[\omega_{\rm gw}t]} \\
    & - \sin{\Big[\omega_{\rm gw}\Big(t - \frac{d_s}{c}\Big(1 + \vu{p} \cdot \vu{n}\Big)\Big) \Big]} \Big\}, 
    \end{split} 
\end{equation} 
where we have now reinserted factors of $c$ and set $t_o=t'$ and $t_e=t'-d_s/c$, and the constant term from integration was dropped. Ultimately, we will be interested in the frequency integral of $R(t)$'s Fourier transform, and that constant term will contribute a delta function centered at $f=0$ to that integrand, which will be outside of the range of integration in Eq. \ref{vamos}.

\subsection{Gravitational-Wave-Detection Statistics}
\label{sec3b}
When collecting data, GW detectors are subject to several sources of noise that complicate the identification of signals from astrophysical sources. The data stream $\mathbf{d}(t)$ received by the detector is the sum of a noise term $\mathbf{n}(t)$ and a (possible) signal term $\mathbf{h}(t)$. Consider $N$ detectors which have the data streams consisting of signals $h_i$ and additive noise $n_i$, $i=1,2,..,N$
\begin{equation}
    d_i(t)=h_i(t)+n_i(t).
\end{equation}
To detect the presence of a signal, we construct a test statistic (TS) from the likelihood ratio:
\begin{equation}
    {\rm TS}=\frac{P(d|\mathcal{H}_a)}{P(d|\mathcal{H}_0)},
\end{equation}
where $\mathcal{H}_a$ and $\mathcal{H}_0$ denote the hypotheses where there is an astrophysical signal and no signal respectively.
Assuming the noise is Gaussian for simplicity \citep[an assumption which allows the noise to be modeled by a one-sided power spectral density and can be checked by measuring the noise of the real detector;][]{2015CQGra..32a5014M} and we are searching for a specific signal $\mathcal{T}_a$, the optimal detection statistic reduces to the well-known matched filter signal-to-noise ratio
\begin{equation}
    \rho= \frac{2\sum_{i=1}^{N}\int_{0}^\infty \frac{\Re(\mathcal{T}_i^*(f)D_i(f))}{S_{1,n,i}(f)}\dd f}{\sqrt{\sum_{i=1}^{N}\int_{0}^\infty \frac{|\mathcal{T}_i(f)|^2}{S_{1,n,i}(f)}\dd f}}
\end{equation}
where $S_{1,n,i}$ are the one-sided power spectral densities of the noises of the detectors (not accounting for the detector response), $D_i$ is the Fourier transform of $d_i$, and $\mathcal{T}_i$ are the signals that would be observed in each detector from the assumed astrophysical signal through the detector responses $R_i(f,p,\hat{n})$, which are functions of direction $\hat{n}$ and polarization $p$
\begin{equation}
    \mathcal{T}_i(f)= \int \sum_p \mathcal{T}_a(f,p,\hat{n})R_i(f,\hat{n},p)e^{-2\pi i f \hat{n}\cdot \vec{x}_i/c}  \dd\hat{n}, 
\end{equation}
where $\vec{x}_i $ are the detector positions.

When the sought astrophysical signal is detected ($D_i(f)=\mathcal{T}_i(f)$), $\rho$ is distributed approximately normally with variance 1 and mean
\begin{equation}
    \label{optimal_SNR}
\Bar{\rho}=2\sqrt{\sum_{i=1}^{N}\int_{0}^\infty \frac{|\mathcal{T}_i(f)|^2}{S_{1,n,i}(f)}\dd f}.
\end{equation}

\subsection{APTA Sensitivity Curve}
\label{sec3c}
With the tools built in Sections \ref{sec3a} and \ref{sec3b}, we may now construct APTA's sensitivity curve for a monochromatic GW, like what a slowly-evolving compact object binary emits.

Consider a source observed for a total time $T$ with observation cadence $1/\delta t$. Then, with the timing residual in Eq. \ref{template} as the time-domain template, whose Fourier transform will be $\mathcal{T}_i(f)$ for the different satellites, and with white noise, for which $S_{1,n}=2\sigma^2 \delta t$, where $\sigma$ is the rms error in the timing residuals, Eq. \ref{optimal_SNR} becomes:
\begin{equation}
\label{vamos}
    \Bar{\rho}^2=\frac{2}{\sigma^2\delta t}\sum_{i=1}^{N}\int_{1/T}^{1/2\delta t} |\mathcal{T}_i(f)|^2 \dd f.
\end{equation}
Without loss of generality, we can orient our coordinates such that $\mathbf{\hat{n}}=\mathbf{\hat{z}}=(0,0,1)$ and $\mathbf{\hat{p}}=(\sin\theta\cos\phi,\sin\theta\sin\phi,\cos\theta).$ Then, $\mathcal{A}_{ij}=h_0(A^+H_{ij}^++A^\times H_{ij}^\times)$, where $h_0$ is an overall amplitude determined by the astrophysical source; $A^{+}=(1+\cos^2\iota)/2$ and $A^\times=\cos\iota$, with $\iota$ being the inclination angle; and $H_{ij}^+=\epsilon_{ij}^+\cos(2\psi)+\epsilon_{ij}^\times\sin(2\psi)$ and $H_{ij}^\times=-\epsilon_{ij}^+\sin(2\psi)+\epsilon_{ij}^\times\cos(2\psi)$, with $\psi$ being the polarization angle, $\epsilon_{ij}^+=\hat{x}_i\hat{x}_j-\hat{y}_i\hat{y}_j$, and $\epsilon_{ij}^\times=\hat{x}_i\hat{y}_j+\hat{y}_i\hat{x}_j$ \citep{2015CQGra..32e5004M}. Hence, the integrand in Eq. \ref{vamos} becomes 
\begin{multline}
    |\mathcal{T}_i(f)|^2 \bigg\rvert_{f>0}=\frac{\delta_T(f-f_{\rm gw})^2}{8 \pi^2 f_{\rm gw}^2}\left(\frac{\hat p^i \hat p^j\mathcal{A}_{ij}}{2(1+\mathbf{\hat p}\cdot\mathbf{\hat n})}\right)^2 \\ \times\left\{1-\cos\left[4\pi f_{\rm gw} \frac{d_s}{c}\cos^2\left(\frac{\theta}{2}\right)\right]\right\}.
\end{multline}where $\delta_T(f)=\sin(\pi f T)/(\pi f)$ is the finite-time delta function. Instead of summing over satellites with different configurations, we determine the average satellite configuration by averaging $|\mathcal{T}_i(f)|^2$ over $\theta$, $\phi$, $\iota$, and $\psi$ and replace the sum by a factor of ${N_s}$, which is the number of satellites. We define $\chi(f_{\rm gw})$ to be the angle-averaged version of the angle-dependent part of $|\mathcal{T}_i(f)|^2$:
\begin{multline}
    \chi(f_{\rm gw})\equiv\int_0^{2\pi} \frac{\dd\psi}{2\pi}\int_0^\pi \frac{\dd\iota \sin\iota}{2}\int_0^{2\pi}\frac{\dd\phi}{2\pi} \int_0^\pi \frac{\dd\theta \sin\theta}{2} \\ \left(\frac{\hat p^i \hat p^j}{2(1+\mathbf{\hat p}\cdot\mathbf{\hat n})}\frac{\mathcal{A}_{ij}}{h_0}\right)^2 \left\{1-\cos\left[4\pi f_{\rm gw} \frac{d_s}{c}\cos^2\left(\frac{\theta}{2}\right)\right]\right\}.
\end{multline}
Next, we can choose a threshold SNR value $\rho_{th}$ to invert Eq. \ref{vamos} and solve for $h_0$. Finally, since we are specializing
to monochromatic binary inspirals, we can convert this result into characteristic strain using $h_c\approx\sqrt{Tf_{\rm gw}}h_0$ \citep{2015CQGra..32a5014M}:

\begin{equation}
\label{sensitivity}
    h_c(f_{\rm gw})=2\pi \sigma\rho_{th}\sqrt{\frac{T \delta t  f_{\rm gw}^3}{N_s \chi(f_{\rm gw})}\left(\int_{1/T}^{1/2\delta t}\delta_T(f-f_{\rm gw})^2 \; \dd f \right)^{-1}}.
\end{equation}

\section{Artificial Precision Timing Array}
\label{sec4}

\begin{figure}
    \centering
    \includegraphics[width=\columnwidth]{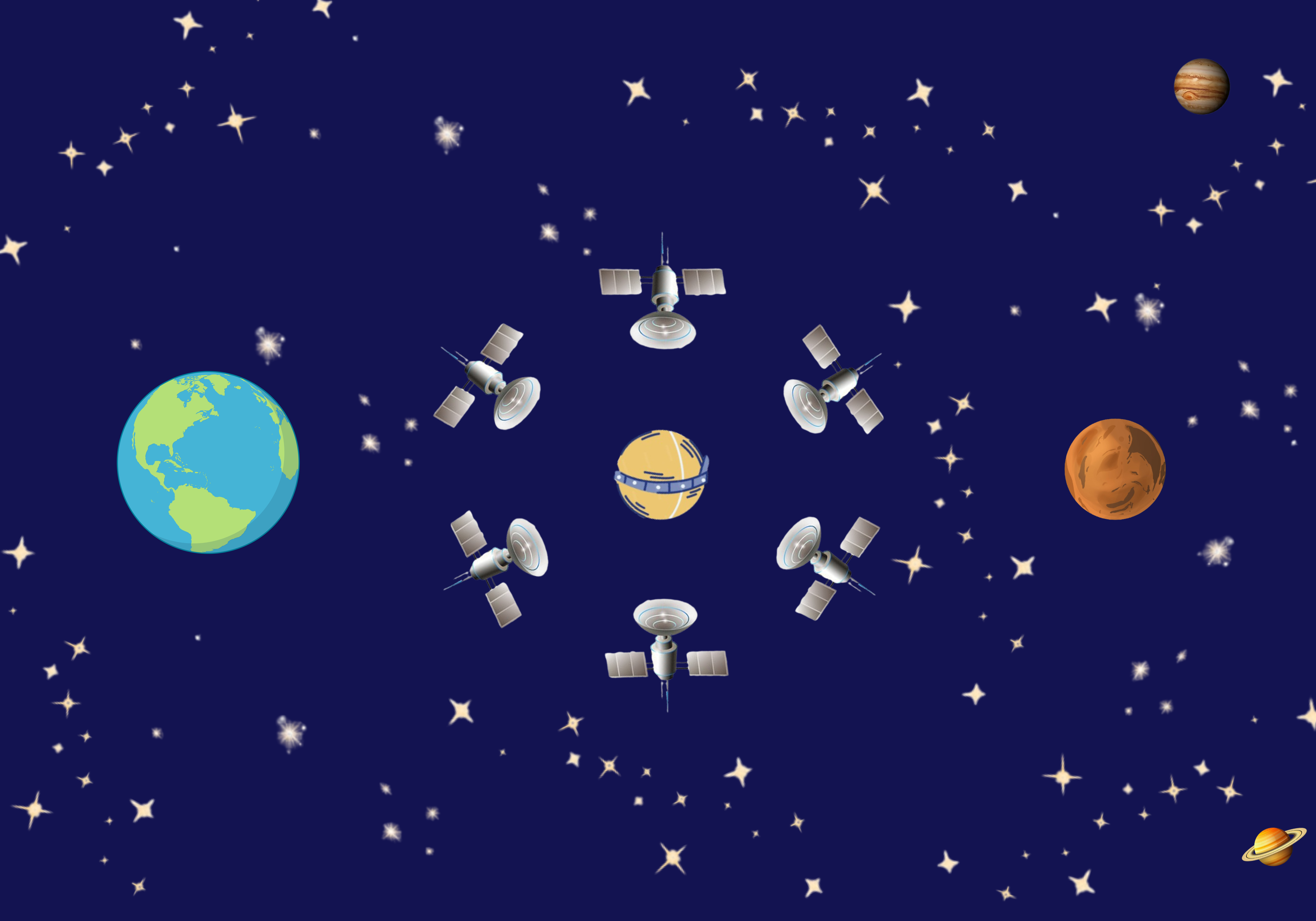}
    \caption{Artist's cartoon illustration of an APTA concept. The detector is composed of an array of satellites communicating with a receiving station, depicted here in the center of the satellite constellation. The satellites, equipped with atomic clocks or comparably accurate time references to allow for precision timing, play the role of the pulsars in this pulsar-timing-like GW detector. Objects are not drawn to scale, and the satellite array's orbit could encircle some solar system planets or the Sun, contrary to what the cartoon figure suggests.} 
    \label{cartoon}
\end{figure}

To construct a GW detector based on PTA principles, we envision distributing periodically-flashing electromagnetic sources in space. These light sources need to have their periodic pulses timed to high precision so that GWs may be observed with them. This can be achieved with an array of satellites carrying atomic clocks or comparably accurate time references, which would play the same role as the astrophysical pulsars in regular PTAs. They could orbit around the Earth or the Sun. Alternatively, rather than Earth-based radio receivers, one could imagine a near-Earth receiving base to avoid atmospheric interference, as illustrated in Fig. \ref{cartoon}. The details of this communication mechanism, including the electromagnetic pulse features and its use in detection, are beyond the scope of this paper, which focuses on clock requirements. To maintain the clock as the primary source of detector noise, the communication noise must be kept lower than the clock's noise.

The results of the model from Section \ref{sec3b} do not depend on the pulsar/satellite communication frequency because $f_{photon}\gg f_{\rm gw}$ for reasonable signal frequency choices \citep{PhysRev.166.1263, 2011A&A...530A..80S}. To observe $0.1$–$10$~Hz GWs, radio waves could do, but visible light would give a pulse-to-GW frequency ratio comparable to that of regular PTAs. X-rays may also become a good choice given recent advances in the generation of high-frequency phase-stable pulses (see e.g. \cite{2022PNAS..11919616Z}). We leave the optimal photon wavelength for APTA's electromagnetic signals as a subject for future investigation.

\subsection{Detector configuration: clock precision and pulsing frequency}
\label{sec4a}

PTAs observing for a total time $T$ with a cadence of $1/\delta t$ are generally sensitive to GW frequencies in the range $(1/T)\lesssim f_{\rm gw}\lesssim(1/\delta t)$ \citep{2015CQGra..32a5014M}. Therefore, accessing the $0.1$–$10$~Hz band can be achieved straightforwardly by searching for GW transients and continuous signals that last at least $\sim10$~s and taking data at least every 0.1~s. 

The characteristic strain sensitivity depends linearly on the error in the PTA timing residuals $\sigma$ as indicated in Eq. \ref{sensitivity}. We will estimate the timing errors needed to conduct significant astrophysics. As we will argue, the sensitivity required is comparable to present and future atomic clock precision. We defer the discussion of additional sources of error to Section \ref{sec5}.

Clocks in development for global navigation satellite systems have reached fractional frequency uncertainties $\xi=\Delta f_{\rm clock}/f_{\rm clock}\sim10^{-15}$ in 1~s of averaging \citep{2021GPSS...25...83S}. More recently, \cite{2025PhRvL.135z3402L} have achieved $\xi=4\times10^{-17}$ at 1~s in a ground-based test of a field-deployable strontium optical lattice clock. State-of-the-art ground-based atomic clocks at present can reach fractional uncertainties $\sim10^{-18}$ in  1~s of averaging and $\sim10^{-19}$–$10^{-20}$ in 1~h \citep{2022Natur.602..420B,2019NaPho..13..714O,2017Sci...358...90C}. In the next few years, we may reach $\xi\sim10^{-20}$ in 1~s averaging with current 3D optical lattice clock technology. In the foreseeable future, use of squeezing and increased atom number may allow $\xi\sim10^{-21}$ \citep{RevModPhys.87.637}. We also remark that intense efforts are underway to miniaturize and ruggedize the technology of strontium optical lattice clocks \citep{poli2014transportable, koller2017transportable, origlia2018towards, takamoto2020test}. Progress has been fast, which makes the prospect of deploying precision optical lattice clocks realistic in the foreseeable future. While reproducing ground-based precision with space-based clocks remains a technological challenge, we shall use as the reference for this study the performance of the best current, near-future, and far-future ground-based atomic clocks. We make this choice since our aim is to provide a proof of concept, to which fundamental physical limitations matter significantly more than present technological ones.

Note that $\xi\propto t_{\rm avg}^{-1/2}$, for $t_{\rm avg}$ the time of averaging. Our averaging time is limited by the cadence, so $\xi_{\delta t}$, the relative clock uncertainty over an averaging time of $\delta t$, is
\begin{equation}
\xi_{\delta t}=\xi_{1 \mathrm{s}}\left(\frac{\delta t}{1 \text{ s}}\right)^{-\frac{1}{2}},
\end{equation}
where $\xi_{1 \mathrm{s}}$ is the fractional uncertainty in 1~s of averaging.
We assume APTA's clocks reach a stable optimal fractional uncertainty $\xi_{\delta t}$ after averaging and use this value to compute the timing error as $\sigma=\xi_{\delta t} \delta t$. 

Regular PTAs are constrained to searching for lower-frequency gravitational waves owing to limited radio telescope access and, ultimately, to the folding time. APTA, on the other hand, can reach much higher frequencies because its precision timing can be carried out effectively continuously. Fundamentally, clocks do not run continuously, but deviations from continuity, as well as the synchronization holdover and the receiver dead time, can be made negligible compared to the period of GWs of interest with the
implementation of dual clocks and interleaved interrogation \citep{Schioppo_2016,2025PhRvL.135z3402L}. Noise in the clocks is effectively suppressed by stabilizing the clock laser to a high-finesse optical cavity, which ensures short-term stability, and near-continuous interrogation of the atomic frequency reference, ensuring long-term stability.

Continued observation can be achieved if the beacon satellites are locked to an observing station, and the folding time may be considerably lowered by the artificial production of a clearly modulated signal. In this case, the cadence $1/\delta t$ is bounded by the frequency at which communication pulses can be stably emitted and received. For communication purposes, an extremely efficient way to transfer the timing precision of optical atomic clocks to the radiofrequency domain is provided by optical frequency comb technology \citep{udem2002optical, diddams2020optical}. It has the potential to generate precision APTA pulses at a carrier frequency of 10 MHz to 100 GHz. The current state of the art for phase-stable X-ray pulses is $10^{-15}$~s \citep{2022PNAS..11919616Z}. Moreover, the periodic light signal that enables precision timing need not come from a pulsing source. For example, high-precision spinning laser sources, cosmic lighthouses, whose low-divergence beams sweep the reception station, could do. For the analysis presented, regardless of what specific mechanism produces the time-reference signal, what matters most is its ability to maintain the stability of the signal's phase and periodicity. The form of the timing signal transmitted can range from a sinusoid to a pulse train and beyond. All of these details will be relevant in future work, but, for this study, all we need is the assumption that a moderate value of $1/\delta t=10$~kHz can be achieved with some of those technology options. Hence, we are focusing on clock uncertainty requirements by utilizing state-of-the-art low values of $\xi$ alongside conservative values of $\delta t$. The same sensitivities could be reached with higher $\xi$ by lowering $\delta t$. The determination of the optimal combination of parameters for constructing APTA is left for subsequent studies.

We return to the discussion of the assumption $\omega_{\rm gw}\tau\ll1$ for APTA (See Section \ref{sec3a}). For APTA, with a clearly modulated signal, there should be no folding-like processes as required for pulsar signals, i.e. there should be no processing time considerably differentiating the pulsing frequency from the observation cadence. Therefore, the inverse cadence $\delta t$ should coincide with $\tau$. Targeting GWs in the 0.1–10~Hz frequency range, current technology is more than enough to ensure $\omega_{\rm gw}\delta t\ll1$. Again, whether smaller values of $\delta t$ would be the optimal choice in practice is beyond the scope of this paper, but we work with Eq. \ref{approx-integral} since it is completely realistic for APTA with respect to technological and astrophysical target requirements.

\subsection{APTA sensitivity curves}
\label{sec4b}
Fundamentally, the sensitivity model involves six free parameters: $\rho_{th}$, $N_s$, $\delta t$, $T$, $\sigma$, and $d_s$. We choose $\rho_{th}=10$ as a standard value for the rest of our study and fiducially pick $N_s=6$. To focus on the dependence on the clock uncertainty as previously discussed, we use $1/\delta t=10$~kHz.  

We assume that all the error in the timing residuals arises from the clock's uncertainty, i.e. $\sigma=\xi_{\delta t}\delta t=\xi_{1 \mathrm{s}}\sqrt{\delta t}$. Motivated by developments in clock technology (see Section \ref{sec4a}), we choose three cases for the clock fractional uncertainty at 1~s of averaging: $10^{-18}$ representing present-day clocks, $10^{-20.5}$ representing near-future clocks, and $10^{-23}$ representing far-future clocks \citep{2022Natur.602..420B,2019NaPho..13..714O,2017Sci...358...90C}.

The model in Section \ref{sec3} concerns the sensitivity of PTAs to monochromatic GW sources, a proxy for slowly evolving binaries. We use $q=1$, $D=1$~Gpc BBHs as discussed in Section \ref{sec2} as example target sources for APTA with total masses of $10^4$, $10^3$, $10^2$, and $10^{-1}$ $\mathrm{M}_\odot$. In the first 10~s after entering the APTA band, the heaviest (and thus fastest-evolving) binary's GW frequency $f$ changes by $\sim1.5\%$, so the monochromatic approximation holds. We initially use $T=10$~s to conservatively cover all four cases.

\begin{figure}
    \centering
    \includegraphics[width=\columnwidth]{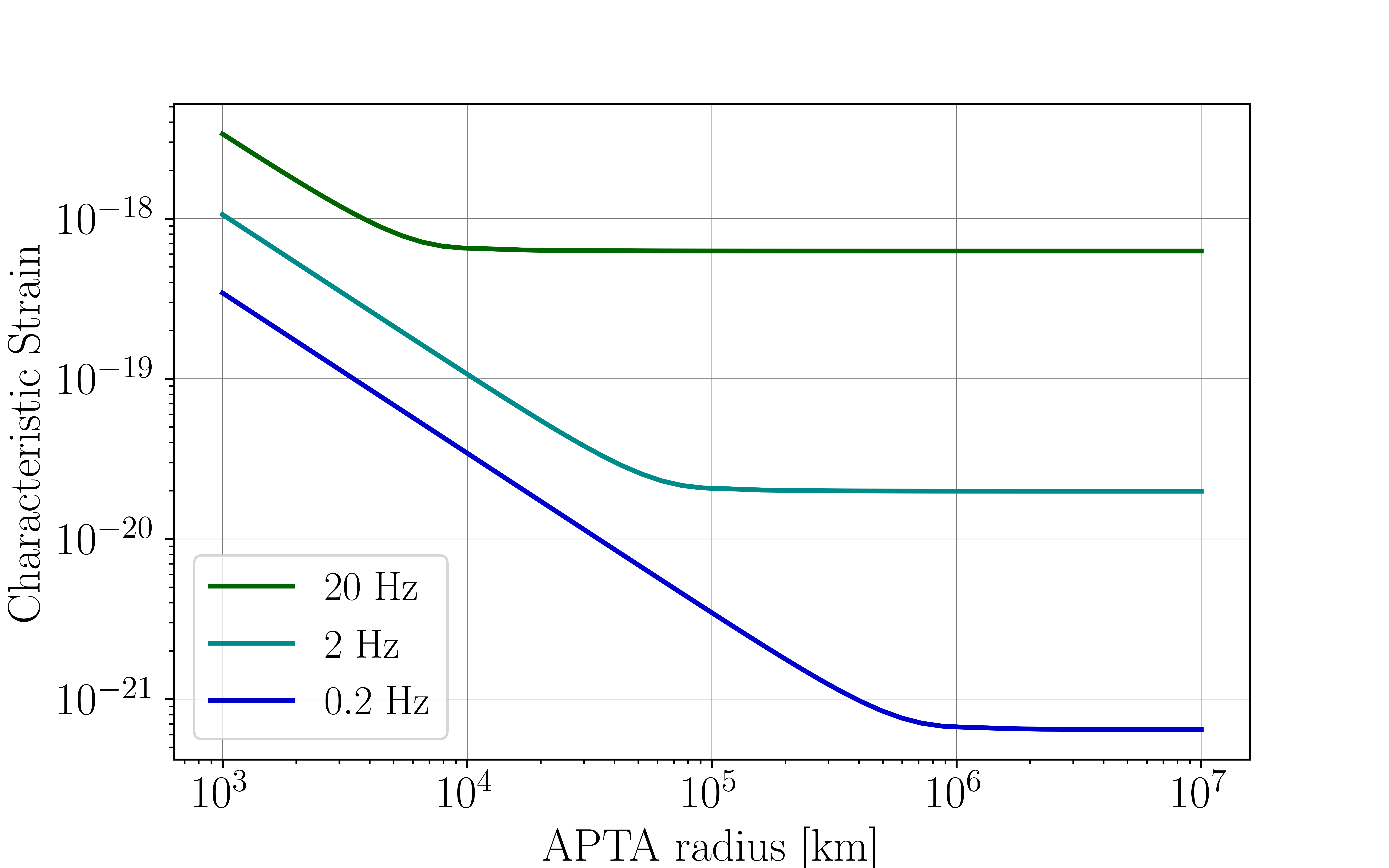}
    \caption{Dimensionless characteristic strain as a function of APTA radius $d_s$ for different GW frequencies. Other input parameters for APTA sensitivity curves are a total observation time $T=10$~s, pulse interval $\delta t=10^{-4}$~s, number of satellites $N_s=6$, and threshold SNR $\rho_{th}=10$.}
    \label{APTAsize}
\end{figure}

The observer-satellite distance affects APTA's sensitivity in Eq. \ref{sensitivity} through the ``pulsar term'' within $\chi(f_{\rm gw})$ (i.e. the term originating from $\cos[\omega_{\rm gw}(t_e-d_s\mathbf{\hat p}\cdot\mathbf{\hat n})]$ in Eq. \ref{template}). The characteristic strain for APTA noise $h_c$ decreases as $\chi$ increases, i.e. the sensitivity is enhanced by the increase of $\chi$. $\chi$ is smaller for $f_{\rm gw}d_s/c=d_s/\lambda_{\rm gw}\ll1$ and increases toward a plateau as this combination of parameters is increased, such that optimal sensitivity is reached if $d_s/\lambda_{\rm gw}$ is large enough, as illustrated in Fig. \ref{APTAsize}. This can be understood as the variation in GW strain sensitivity becoming negligible when the array's size is made large enough compared to the GW wavelength—this approximation is good and widely used in nHz GW searches with regular PTAs. To reach that optimal sensitivity in 0.1–10~Hz GW searches, a LISA-sized APTA with $d_s=d_{\rm LISA}\equiv2.5\times 10^6$ km is sufficient \citep{LISA:2017pwj}, so we will be using this value as an important reference.

\begin{figure}
    \centering
    \includegraphics[width=\columnwidth]{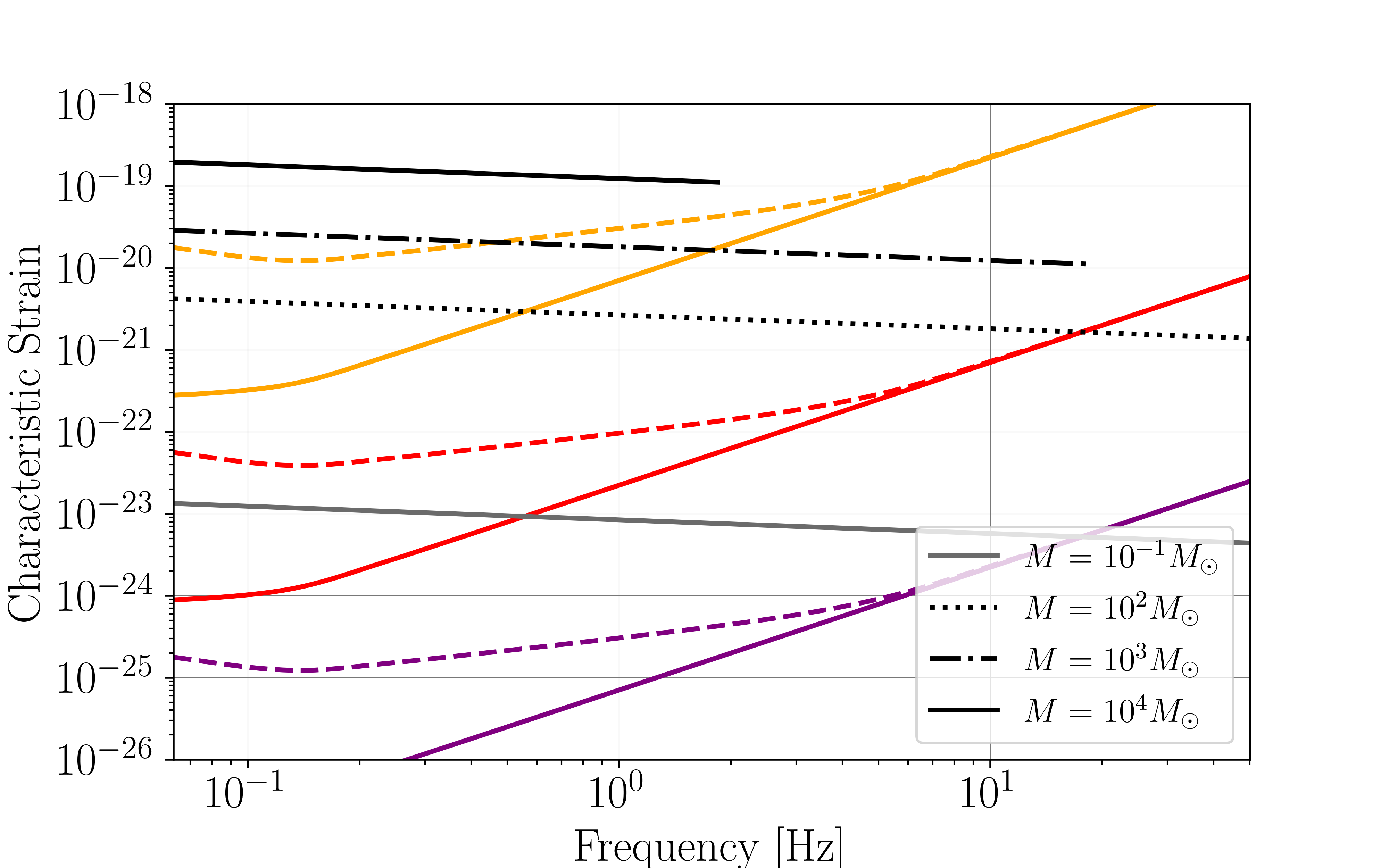}
    \caption{Dimensionless characteristic GW strain as a function of frequency  for different APTA configurations and binary black hole total masses. The colored lines are APTA sensitivity curves with $T=10$~s, $\delta t=10^{-4}$~s, $N_s=6$, and $\rho_{th}=10$; solid ones have $d_s=d_{\rm LISA}$ while dashed ones have $d_s=d_{\rm LISA}/100$; yellow indicates $\xi_{1 \mathrm{s}}=10^{-18}$, red, $\xi_{1 \mathrm{s}}=10^{-20.5}$, and purple, $\xi_{1 \mathrm{s}}=10^{-23}$. The black and gray lines represent binary black holes with mass ratio $q=1$ at a distance $D=1$~Gpc, with masses as indicated. The solid and dash-dotted black lines representing the binaries with total masses $M=10^4$ $\mathrm{M}_\odot$ and $M=10^3$ $\mathrm{M}_\odot$ end at their respective merger frequencies (see Eq. \ref{eq:mergefreq}).} 
    \label{fig:BBH}
\end{figure}

In Fig. \ref{fig:BBH}, we display APTA sensitivity curves alongside example GW strains of merging intermediate-mass BBHs and early inspiraling stellar-mass and primordial BBHs. 
With current ground-based atomic clock technology, represented by $\xi_{1 \mathrm{s}}=10^{-18}$ \citep{2022Natur.602..420B,2019NaPho..13..714O,2017Sci...358...90C}, APTA could already observe $10^3$ $\mathrm{M}_\odot$ and $10^4$ $\mathrm{M}_\odot$ BBHs at or close to merger at~Gpc distances, and, for $d_s=d_{\rm LISA}$, see a part of the evolution of 100 $\mathrm{M}_\odot$ binary inspirals before they enter the LIGO band. With technological progress toward values of $\xi_{1 \mathrm{s}}$ around $10^{-20.5}$, the 0.1 $\mathrm{M}_\odot$ PBH binary would become observable by the larger APTA, and better clocks could make the detection of lighter and lighter PBHs possible.

\begin{figure*}
    \centering
    \subfigure{
        \includegraphics[width=\columnwidth]{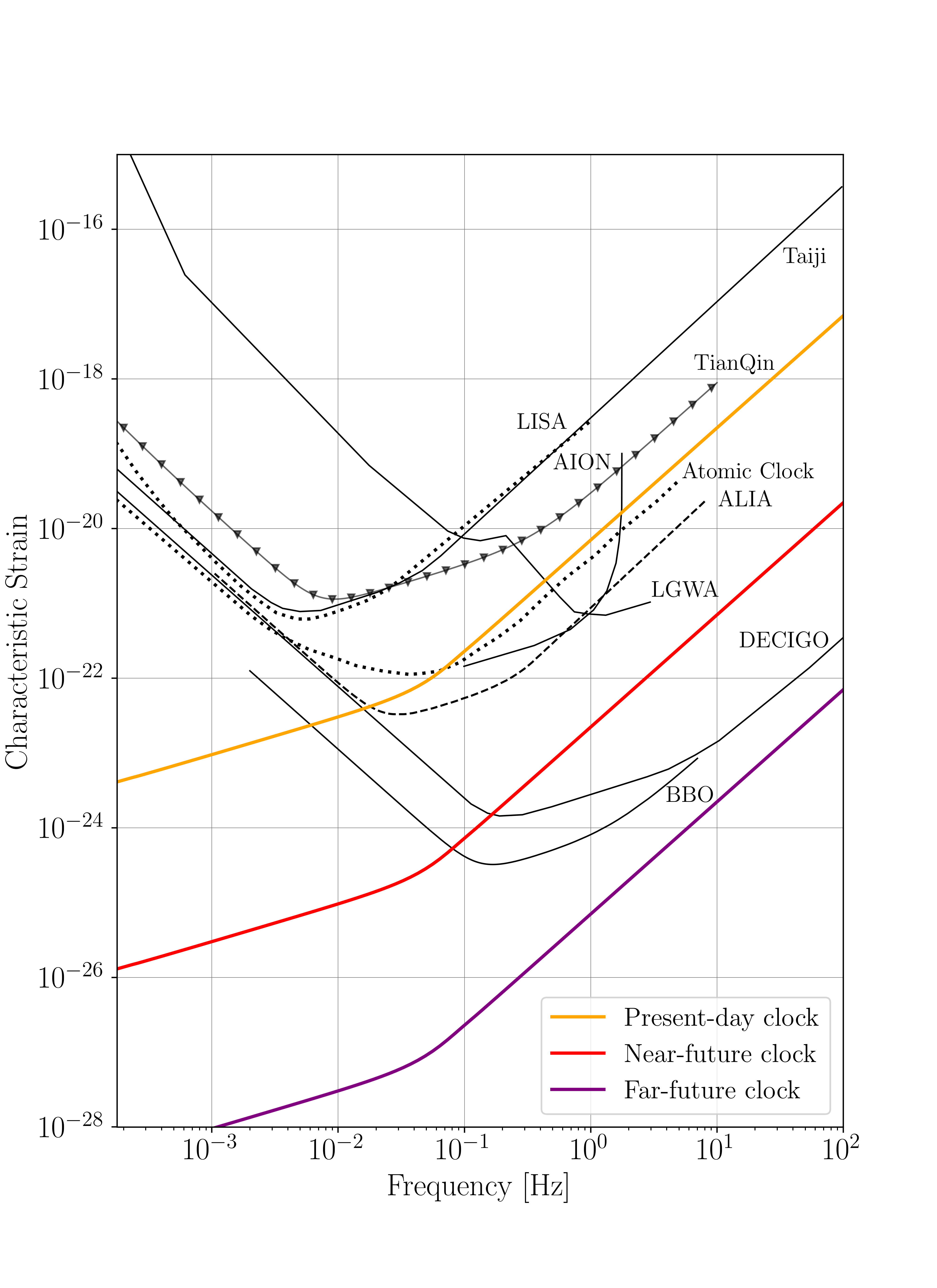}
    }
    \subfigure{
        \includegraphics[width=\columnwidth]{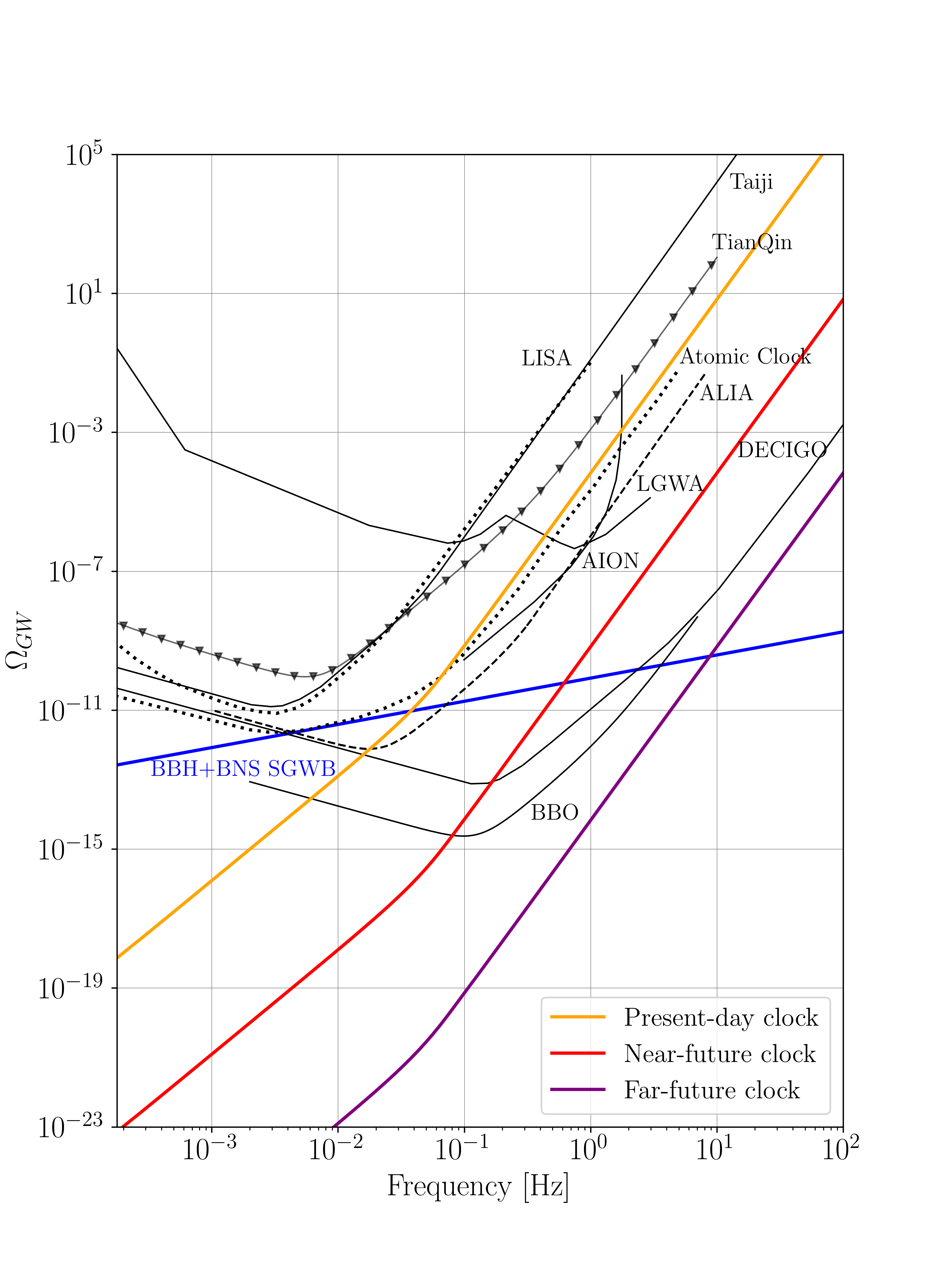}
    }
    \caption{Dimensionless characteristic GW strain as a function of frequency for different detectors. We display APTA for $[T,\delta t,\rho_{th},N_s,d_s]=[10^4\mathrm{\;s},10^{-4} \mathrm{\;s},10,6,2.5\times10^6\;\mathrm{km}]$ with different colors indicating different values of $\xi_{1 \mathrm{s}}$ as discussed in the text: $10^{-18}$ for present-day ground-based clocks, $10^{-20.5}$ for near-future clocks, and $10^{-23}$ for far-future clocks. Other detectors, shown in black, are Taiji \citep{2020IJMPA..3550075R}, TianQin \citep{Luo_2016}, LISA \citep{LISA:2017pwj}, AION (in its most sensitive configuration, AION-1km) \citep{Badurina:2019hst}, an atomic-clock-based technique \citep{PhysRevD.94.124043,Sedda_2020}, ALIA \citep{Bender:2004vw,bender_begelman_2005}, LGWA \citep{ajith2024lunargravitationalwaveantennamission}, DECIGO \citep{Kawamura:2006up}, and BBO \citep{phinney2004big,Harry_2006}. In the right-hand-side plot, we additionally display in blue the LIGO-Virgo-KAGRA prediction for the SGWB produced by BBHs and BNSs in our universe \citep{SGWB_LVK}.} 
    \label{APTA-friends}
\end{figure*}

In Fig. \ref{APTA-friends}, we showcase the sensitivity of the LISA-sized APTA alongside many future detectors in terms of the characteristic strain $h_c$ and the gravitational-wave energy density parameter per logarithmic frequency $\Omega_{\rm GW}(f)=2\pi^2f^2h_c(f)^2/3H_0^2$ \citep{2015CQGra..32a5014M}, where $H_0=67.4$ km/s/Mpc is the Hubble constant \citep{Planck:2018vyg}, related to the current rate of expansion of the universe. We focus on proposed detectors with the potential of observing astrophysical sources in the decihertz band. For this figure, $T$ is raised to $10^4$~s since precision timing can only be used to observe GWs with $f\gtrsim1/T$. While, for some sources, this is inconsistent with our assumption of monochromatic waves, it is still a reasonable timescale for the observation of compact binary inspirals and other sources in APTA's band, and a realistic APTA detector should certainly be modeled and constructed to follow GWs whose frequencies evolve in time.

With current ground-based clock technology \citep{2022Natur.602..420B,2019NaPho..13..714O,2017Sci...358...90C}, APTA can achieve better sensitivity than Taiji, LISA, and TianQin for all frequencies displayed in Fig. \ref{APTA-friends} and outperform AION, the atomic clock concept, ALIA, LGWA, DECIGO, and BBO in some frequency ranges. Progress toward lower clock uncertainties would allow APTA to surpass all of its existing and proposed peers for all frequencies.

In the right panel of Fig. \ref{APTA-friends}, we also include the LIGO-Virgo-KAGRA prediction for the stochastic gravitational-wave background (SGWB) produced by BBHs and binary neutron stars (BNSs) \citep{SGWB_LVK}. Inspiraling compact binaries are known to create an SGWB spectrum $\Omega_{\rm SGWB}(f)\propto f^{2/3}$ \citep{phinney:2001di}. We ignore deviations from this due to close-to-merger behavior and use the $2/3$ spectral index across all frequencies. Although we have not modeled APTA's sensitivity to an SGWB, but solely to localized sources, the comparison between our sensitivity curves and the predicted SGWB signal remains informative because we only expect $\mathcal{O}(1)$ factors to appear following a more detailed analysis of a stochastic-background signal.

\section{Towards a more realistic model of APTA}
\label{sec5}

This study’s main goal is to estimate the fundamental capabilities of detecting GWs with artificial ``pulsar” timing. In a more realistic model of APTA, several additional effects must be taken into account. Although detailed calculation of these effects is beyond our present scope, we discuss some of them below.

Shapiro time delay, the change in the travel time of light between two points due to the presence of massive bodies \citep{PhysRevLett.13.789}, from various solar system objects will introduce delays of order $2GM/c^3$, e.g. $\sim10^{-5}$~s for the Sun, $\sim10^{-8}$~s for Jupiter, and $\sim10^{-13}$~s for the Moon. Since the main asteroid belt has an estimated total mass of $18\times10^{-10}$ $\mathrm{M}_\odot$ \citep{KRASINSKY200298}, asteroids may also contribute with delays $\sim10^{-14}$ s. These known delays will be included in the timing model. The gravity gradient noise of asteroids, on the other hand, only contributes marginally at much lower frequencies ($10^{-9}$–$10^{-5}$~Hz) \citep{PhysRevD.105.103018,Sullivan_2022}, so should not affect APTA.

In this study, we have utilized monochromatic binaries as example target sources for APTA. In reality, sources evolve in time. By not integrating the signal over the entire frequency band and not using any information about the signal’s time dependence, we underestimate the sensitivity of APTA. In future analyses, accounting for the evolution of signals in time will allow for the determination of how much better APTA’s sensitivity could be in an actual implementation.

In reality, APTA's sensitivity curves in Figs. \ref{fig:BBH} and \ref{APTA-friends} should lose sensitivity at very low frequencies. Our simple models do not predict this because we have assumed all noise is clock-timing noise with a fixed uncertainty.  A more realistic model would include error in the timing residual $\sigma$, cadence $1/\delta t$, and observing time $T$, which can vary from satellite to satellite and observation to observation \citep{2015CQGra..32e5004M}, as well as various unresolved astrophysical sources at low frequencies \citep{1990ApJ...360...75H}. This is a subject of future study.

Because of APTA's small solar-system scale, some components of traditional PTA models, including parallax \citep{Deng_2011} as well as Einstein and Roemer delays (related to the use of a coordinate system centered in the solar system barycenter \citep{2006MNRAS.369..655H}), are not relevant. 
The solar wind \citep{2006LRSP....3....1M} contributes to the dispersion measure (DM), i.e. the integrated column density of free electrons between an observer and a light source. While accounting for the effects of the time-dependent DM on the ToAs is not straightforward \citep{NANOGrav:2023ctt}, this problem is easier for APTA than for PTAs, since contributions only come from the solar system and not from the interstellar medium.

After setting the satellites in orbit, their positions should be tracked with sufficient precision to account for their motion and added time delays. Additionally, it will be important to track disruptions to the network (e.g. a satellite being perturbed by a planet or colliding with an asteroid). Disruptions to one or a few elements will not necessarily prevent GW detection since the signals detected by the other satellites remain correlated. If further analyses reveal that this is an important concern, the detector can be designed with a redundantly large number of satellites.

To first approximation, clock synchronization should not play an important role in the functioning of APTA. Surely, regular PTAs are not synchronized clocks, and have already been proven successful by relying on the long-term timing precision of the individual pulsars. Nevertheless, characterizing the stationarity of the clock signal is important, especially in space, which is not a controlled environment. Besides the internal characteristics of the clocks, the environment also needs to be monitored. The fundamental capacity to compare satellite clocks to a reference and to each other at the reception station can also reveal unmodeled glitches of the space environment. In general, significant time precision losses may happen in the process of time transfer (see e.g. \cite{2023Natur.618..721C}). Preserving, transferring, and utilizing the satellite clocks' precision is also mission-critical for success. Therefore, the generation of the electromagnetic pulses and their locking to the clock should preserve the accuracy of the time reference. This can be done with state-of-the-art frequency comb technology \citep{udem2002optical, diddams2020optical}. Additionally, the pulses' phase information and intensity must not degrade sufficiently during their travel as to compromise the recovery of the timing information. The multi-beam detection scheme at the receiver location must also preserve the mission-critical relative timing information. Each of these interwoven problems will require its own individual technical analysis as the required clock accuracy only sets the performance specifications of the downstream systems engineering. 

\section{Conclusion}
\label{sec6}

We have presented a proof-of-principle study into the use of an artificial precision timing array to detect 0.1–10~Hz GWs, focusing on the required time reference precision. APTA would consist of a few satellites equipped with precision clocks orbiting and communicating preferably with a space-based reception station so the atmosphere is not in the path. The desired frequency band can be reached by targeting astrophysical signals at least 10~s long and ensuring that the beacons' pulses are received at least every 0.1 s. APTA would not need to be larger than LISA and already performs significantly well with $1\%$ of LISA's size.

Compared to traditional PTAs, APTA could enable better GW source localization, mainly because of the high precision in the knowledge of its satellites' positions and lower clock uncertainty. Modern solar system ephemerides such as DE440 are advanced enough that they produce positional errors of less than 1 km for most celestial bodies across the solar system \citep{park_jpl_2021}. This hints at great prospects for APTA compared to traditional PTAs, which have pulsar positional uncertainties $\gtrsim100$ pc. The effect of this improved precision has been studied numerically for the pulsar timing case, where the angular uncertainty area decreased by almost 5 orders of magnitude as precision increased from 100 pc to 0.01 pc. In addition, when compared to typical PTA timing uncertainties of $\lesssim100$ ns, APTA's higher-precision clocks could considerably reduce the angular uncertainty area \citep{kato_precision_2023}.

APTA's sensitivity is fundamentally limited by $\sigma$, the error in the timing residuals. With the current state-of-the-art ground-based atomic clock technology \citep{2022Natur.602..420B,2019NaPho..13..714O,2017Sci...358...90C}, APTA could achieve better dHz sensitivity than LISA, TianQin, and Taiji, allowing for the detection of $10^3$–$10^4$ $\mathrm{M}_\odot$ black hole mergers and the early inspirals of some ground-based GW sources. With future clock technology available \citep{RevModPhys.87.637}, APTA could compete with and eventually surpass the sensitivity of other proposed detectors for the 0.1–10~Hz frequency range without the same challenges in high-precision interferometry they face. Then, it could detect most astrophysical sources in this band. These exciting prospects motivate the continued development of precision clock technology and its application to GW detection.

While focusing on the accuracy of APTA's time reference, we have only discussed briefly the effects of additional noise sources, technical challenges, and technical developments required. This will require more detailed future work. Nevertheless, with conservative choices for measurement cadence $\delta t$, we have shown that high sensitivity can be still be achieved, raising optimism for the APTA concept. With this and future papers laying the foundation, we propose APTA to observe uncharted territory of the gravitational-wave sky and gain precious new insight into high-energy astrophysics and cosmology.

\section*{Acknowledgements}

The authors thank Columbia University in the City of New York and Stanford University for their generous support. LMBA is grateful for the Columbia Undergraduate Scholars Program Summer Enhancement Fellowship, the Columbia Center for Career Education Summer Funding Program, and the support from Coordenação de Aperfeiçoamento de Pessoal de Nível Superior (CAPES). AS is grateful for the support of the Stanford University Physics Department Fellowship, the National Science Foundation Graduate Research Fellowship, and a Giddings Fellowship to the Kavli Institute for Particle Astrophysics and Cosmology at Stanford. DV is grateful to The Scientific and Technological Research Council of T\"urkiye for their support through the grants 123C484 and 123C213; and to the European Commission through the MSCA COFUND Programme with the Grant Agreement No. 101081645. Especially, we would like to thank Peter Bender, Rainer Weiss, Stephen Taylor, Gianluca Giavitto, and Marek Kowalski for their ideas, encouragement, help, and suggestions.

\section*{Data Availability}
The data underlying this study will be made available upon reasonable request to the author.
\bibliographystyle{mnras}
\bibliography{Refs}
\label{lastpage}
\end{document}